\documentclass[a4paper,onecolumn,11pt,accepted=2026-01-25]{quantumarticle}
\pdfoutput=1
\usepackage[utf8]{inputenc}
\usepackage[english]{babel}
\usepackage[T1]{fontenc}

\usepackage{tikz}
\usepackage{lipsum}

\usepackage{amsthm,amsfonts,amsmath,amssymb,bm,verbatim}
\usepackage{chngcntr,colortbl,epsfig,graphicx,mathtools,xr-hyper,hyperref,multirow,tabularx,tikz,xcolor}
\usepackage{arydshln}
\usepackage[misc]{ifsym}
\usepackage[all]{xy}
\usepackage{algorithm,setspace}
\usepackage[noend]{algpseudocode}
\usepackage{multirow}

\usepackage[numbers]{natbib}

\newtheorem{definition}{Definition}{\bfseries}{}

\let\oldv=\v
\renewcommand{\v}[1]{{\bm #1}}

\newtheorem{myRemark}[definition]{Remark}{\bfseries}{\itshape}
\newtheorem{myClaim}[definition]{Claim}{\bfseries}{\itshape}
\newtheorem{myProposition}[definition]{Proposition}{\bfseries}{\itshape}
{\bfseries}{\itshape}
\newtheorem{myAssumption}[definition]{Assumption}{\bfseries}{\itshape}

\DeclarePairedDelimiter\ket{\lvert}{\rangle}
\DeclarePairedDelimiterX\braket[2]{\langle}{\rangle}{#1 \delimsize\vert #2}

\newcommand{\algrule}{\Statex\par\vskip2pt\hrule\par\vskip-2pt}
\algrenewcommand\algorithmicrequire{\textbf{Input:}}
\algrenewcommand\algorithmicensure{\textbf{Output:}}

\graphicspath{{../figures/}{./figures/}{../../figures/}} 

\begin{document}

\title{Quantum Circuit Optimization by Graph Coloring}

\author{Hochang Lee}
\orcid{0009-0006-2387-2364}

\author{Kyung Chul Jeong}
\orcid{0000-0001-7988-6761}

\author{Panjin Kim}
\email{pansics@nsr.re.kr}
\orcid{0000-0002-2657-311X}
\affiliation{The Affiliated Institute of ETRI, Daejeon 34044, Korea}

\maketitle

\begin{abstract}
This work shows that minimizing the depth of a quantum circuit composed of commuting operations reduces to a vertex coloring problem on an appropriately constructed graph, where gates correspond to vertices and edges encode non-parallelizability.
The reduction leads to algorithms for circuit optimization by adopting any vertex coloring solver as an optimization backend.
The approach is validated by numerical experiments as well as applications to known quantum circuits, including finite field multiplication and QFT-based addition.
\end{abstract}

\section{Introduction}\label{sec:1}



Quantum circuits composed of finitely many unitary operations that mutually commute arise in several computational settings, including instantaneous quantum polynomial\;\cite{IQP,IQP2}, commuting local Hamiltonians\;\cite{CLH,toric}, and oracle constructions in quantum algorithms\;\cite{MulMaslov,draper}.
Throughout this work, we refer to such circuits as commuting circuits.
Because the underlying unitary operations mutually commute, such circuits are naturally amenable to parallelization, i.e., certain operations can be executed in parallel whenever the hardware supports it.
Hardware-dependent constraints therefore determine which operations can be executed concurrently, and consequently lead to the notion of \textit{depth} of a circuit.

In circuit synthesis, a function can generally be implemented by multiple constructions, and selecting one that minimizes resource usage is a central consideration in circuit design\;\cite{SM13}.
Under a fixed number of qubits, a design with smaller depth is typically preferred, as it reflects better utilization of the available parallelism. 
For this reason, among others, optimizing the depth of a circuit is an active area of research\;\cite{SM13,nam18,BAA+20,HS22_TDepth,AMM14}.

This study establishes a formal connection between depth optimization of commuting circuits and a well-known NP-hard problem in graph theory. 
The results are summarized as follows:

\begin{itemize}
  \item[$\bullet$] Depth optimization of a commuting circuit is shown to be polynomial time reducible (in the number of operations) to solving the vertex coloring problem. The opposite direction is also proven to hold, meaning that the two problems are equivalent.
  \item[$\bullet$] The reduction implies \textit{any algorithm} for the vertex coloring can be utilized to optimize a commuting circuit.
  \item[$\bullet$] Two concrete algorithms are numerically tested to validate the method.
  \item[$\bullet$] The applicability is demonstrated by optimizing circuits for finite field multiplications and additions by quantum Fourier transform (QFT).
  The first application exhibits the improvement upon Maslov et al.'s design\;\cite{MulMaslov} and the second one gives tradeoff options for Draper's quantum addition\;\cite{draper}.
\end{itemize}


The paper is organized as follows.
Section\;\ref{sec:preliminary} introduces the necessary preliminaries.
Section\;\ref{sec:main} presents the main reduction and its proofs, followed by algorithms in Section\;\ref{sec:algorithms}.
Section\;\ref{sec:applications} discusses applications of the method, and Section\;\ref{sec:summary} concludes the paper.




\section{Preliminaries}\label{sec:preliminary}
This section introduces necessary definitions, known facts, and related works.
Readers are assumed to be familiar with the basics of quantum information\;\cite{bookchuang}.

\subsection{Parallel Execution of Commuting Operations}\label{sec:motivation}
The purpose of this subsection is to bring out the motivation. 
Consider a Toffoli gate denoted by $C\!X_{ijk}$ where $i$, $j$ are indices of the control bits and $k$ is the index of the target bit.
Let $\v v_1$ be an ordered set, $\v v_1 = (C\!X_{134}, C\!X_{235})$ and consider a quantum circuit implementing gates in $\v v_1$.
Notice that the two gates share the same control bit (indexed by 3), and we assume a quantum machine cannot execute these gates simultaneously.
Therefore in this case we count the depth of $\v v_1$ as $2$.
(Definition of the depth is given below.)

Now let $\v v_2 = (C\!X_{134}, C\!X_{235}, C\!X_{167}, C\!X_{268})$.
The depth of $\v v_2$ is 3 since $C\!X_{235}$ and $C\!X_{167}$ do not share any control or target bit, thus the two gates are simultaneously executable by the machine (assumed).
There is a way to further decrease the depth of the circuit.
Notice that the gates in $\v v_2$ are pair-wise commuting; any permutation of the four gates leads to the same behavior.
If one picks an order $\v v_2' = (C\!X_{134}, C\!X_{268}, C\!X_{235}, C\!X_{167})$, and then the depth becomes $2$.


\subsection{Gate Ordering Problem}\label{sec:gate-ordering}
\begin{definition}[Commuting operations]
	Two unitary operations written by matrices $v,w \in \mathbb{C}^{z \times z}, z\in \mathbb{Z}^+$ are \textit{commuting} if and only if $v w = w v$.
\end{definition}

\begin{definition}[Depth of a circuit]
	Given a hardware system, operations are parallelizable if and only if they can be simultaneously executed on the system.
	A layer is the maximal set of operations such that the operations are parallelizable.
	Depth of a circuit is the number of consecutive layers to run the circuit.
\end{definition} 


Be reminded that depth is defined in terms of operations, not mere elementary gates.
For example in a modular exponentiation circuit for modulus $N$ (that appears in factoring or discrete logarithm algorithms), the modular multiplication takes place $O(\log N)$ times, thus if we count the multiplication-depth it would read $O(\log N)$.
However, the multiplication can be further decomposed into $O(\log N)$ additions, in which each addition takes another $O(\log N)$ elementary operations, resulting in $O((\log N)^3)$ depth of elementary gates for the exponentiation.
Notice in this example that modular multiplications are all pair-wise commuting whereas not all elementary gates are commuting during the exponentiation.

From now on, the term \textit{gate} will often be used instead of operation.
Let $\v v$ be a sequence of gates where the gates are pair-wise commuting. 
The problem we want to solve is defined as follows:

\begin{definition}[Gate ordering]
	For a given circuit $\v v$ consisting of commuting gates, a gate ordering problem is to reorder the gates in $\v v$ such that the depth is minimized.
	%
\end{definition}
%
Here we do not consider cancelling or replacement of a subcircuit\;\cite{maslov21}, for example, applying the same CNOT gate twice in a row is equivalent to doing nothing, but such optimization is not a point of interest.

\subsection{Graph coloring}\label{sec:graph-coloring}
A graph is an ordered set $G=(V,E)$, where $V$ is a set of vertices $v_1, \ldots, v_{|V|}$ and $E$ is a set of edges denoted by $e_{ij}$ connecting two vertices $v_i$ and $v_j$.
Only simple and undirected graphs are considered in this work\;\cite{bondy2008graph}.

A graph coloring of $G=(V,E)$ is a function $c:V \rightarrow C$, where $C$ is a set of colors.
Without loss of generality, we will use a convention that the colors are drawn incrementally from the natural numbers, $C= \{1,2,3,\ldots\} \subseteq \mathbb{N}$.



\begin{definition}[Proper coloring]
	A proper coloring $c$ of $G$ is a coloring such that $c(v_i) \ne c(v_j)$ for all $e_{ij} \in E$.
	\label{def:colored-graph}
\end{definition}
Notice that $e_{ij} \notin E$ does not mean $c(v_i) = c(v_j)$.
Only the proper coloring is considered in this work, thus `proper' will be dropped hereafter.

A colored graph of $G=(V,E)$ by a coloring $c$ is denoted by $\widetilde{G} = (V,E,c)$.
Define $\mathbf{Col}(G)$ as the set of colored graphs of $G$.

\begin{definition}[Vertex coloring problem]
	Given a graph $G=(V,E)$, a vertex coloring problem is to find a colored graph $(V,E,c) \in \mathbf{Col}(G)$ such that $| \{ c(v): v\in V \} |$ is minimized.
	\label{def:graph-coloring}
\end{definition}

The term graph coloring frequently refers to vertex coloring throughout the paper. 
Chromatic number of a graph is the smallest number that is required to color the graph.



\subsection{Related works}
The optimization of a logic circuit has long been studied.
The minimum circuit size problem asks if a Boolean function is computable by a circuit of limited size, and several papers of evidence suggest that the problem might be NP-hard\;\cite{KC00,AHM+08,SIR18,RBI+20,Hir22}.
In quantum computing, recent works show that optimizing a quantum circuit in some occasions is NP-hard\;\cite{braid,WA24}, and specifically in the paper by Wetering and Amy\;\cite{WA24}, wide range of optimization targets have been considered including $T$-count, $T$-depth, Toffoli-count, Toffoli-depth, CNOT-count and so on, proving that the quantum circuit optimization is generally hard.
Along the line of study, one of the conclusions drawn from this work is that the depth optimization of a commuting quantum circuit is NP-hard, implying that this small class of quantum circuits is already hard to optimize.

Circuit optimization is one of the active research areas in the quantum computing community.
On the one hand, it is to ask theoretically how small a circuit can be, but on the other hand, finding better optimization is closely related to overcoming or mitigating noise problems\;\cite{preskill18}.
The aspect of optimization varies, for example $T$-depth optimization of Clifford+$T$ circuits\;\cite{amy14}, pattern matching\;\cite{maslov05}, local synthesis\;\cite{soeken10}, or design automation\;\cite{MQT}.
This work pursues an optimization by the reordering of gates, in which the goal and the approach are different from previous works.
The closest notion in the literature is the commutation analysis (built-in function of Qiskit\;\cite{qiskit2024}) whose purpose is to find cancellations or mergers, or to lower the overhead accounting for a hardware topology.
We do not consider cancellations or so, but are interested in how much parallelization is achievable solely by reordering.

There exist works on quantum circuit optimization utilizing coloring in graph theory, by looking qubits as vertices.
For example, edge coloring of a graph has been used to optimize the depth of Clifford circuits\;\cite{MZ22}, or binary field inversion\;\cite{mulGNB}.
Such an approach is natural in the sense that once wires (qubits) are viewed as vertices and gates are considered as edges interconnecting the wires, the result of edge coloring tells which gates are parallelizable because two edges being connected to the same vertex implies they share the same qubit as the reference.
Moreover, in cases where gates operating on more than two qubits are used, the approach can be generalized by considering hypergraphs where an edge can connect more than two vertices.
For example in\;\cite{HOH21}, the lower bound on the depth of the single optimization layer is given in terms of the chromatic index of the hypergraph\;\cite{bookBergeHypergraph}.

Graph coloring has been used to optimally group Pauli strings to minimize the number of measurements in variational quantum eigensolver\;\cite{JGM19,VYI20,HI20,GAD20,VQE}.
Given a set of Pauli strings, one tries to partition them into subsets such that all strings within each subset commute, while minimizing the number of subsets.
The strings in each subset are simultaneously diagonalizable, allowing simultaneous measurement.
This problem is reduced to solving graph coloring where the edges are drawn if two strings (vertices) are not commuting.
The optimal partitioning of a set of Pauli strings is also considered in reducing the depth of a circuit for the time evolution of Sachdev-Ye-Kitaev model\;\cite{AJS24}.
The authors successfully reduce the depth of the circuit by using the graph coloring algorithm.

We extend the idea by considering a circuit \textit{consisting only of commuting operations}. 
The optimization approach developed in this paper uses an algorithm for graph coloring as the core routine.
Any graph coloring algorithm can be used, for example including ones from previous works\;\cite{WP67,ColDSatur,Lei79,AS06}, and it will soon be shown that the output of the graph coloring is directly related with the optimized circuit. 


\section{Reduction from gate ordering to graph coloring}\label{sec:main}
The gate ordering problem is reduced to the graph coloring problem in this section.
The opposite direction is also given in Appendix\;\ref{sec:reduction-other-way}, showing that the two problems are equivalent.
Since the graph coloring problem is in NP-hard~\cite{bondy2008graph}, it implies optimizing a commuting circuit is also hard.


\subsection{Transformations}\label{sec:circuit_to_graph}
Two transformations are introduced before we give the reduction.
One is from a circuit to a graph, and the other is from a colored graph to a circuit.

\subsubsection{Circuit to graph}
Let $\v v$ be an ordered set of commuting gates, and let $\cal V$ be a multiset of gates in $\v v$ (where identical elements may appear multiple times).
One can think of $\v v$ as a circuit and $\cal V$ as a collection of gates in the circuit.
Given $\v v$, construct a graph $G_{\cal V}$ of which \textit{vertices are gates} by the following algorithm:

\begin{algorithm}[H]
	\small
	\caption{$\mathsf{ConstructGraph}$}
	\begin{algorithmic}[1]
		\Require $\v v = (v_{1}, v_{2}, \ldots )$
		\Ensure $G_{\cal V} = (V, E)$
		\algrule \vspace{2pt}
		\State $V \gets \{v_{1}, v_{2}, \ldots \}$, $E \gets \{~\}$
		\For {$i = 1$ \textbf{to} $|V|$}
		\For {$j = 1$ \textbf{to} $i-1$}
		\If {$v_{i}$, $v_{j}$ are not simultaneously executable}
		\State $E \gets E \cup \{e_{ij}\}$
		\EndIf
		\EndFor
		\EndFor
		\State \Return $(V, E)$
	\end{algorithmic}
	\label{alg:construct_graph}
\end{algorithm}
\noindent It is a fairly natural construction.
Gates that cannot be executed simultaneously correspond to \textit{connected} vertices that cannot be identically colored.
Therefore one first draws vertices corresponding to gates in $\cal V$, and then draws edges between vertices depending on the parallelizability of the gates.
About the time complexity, it checks the parallelizability of two gates $O(|{\cal V}|^{2})$ times in total.

Notice that any permutation of $\v v$ results in the same graph by Algorithm\;\ref{alg:construct_graph}.
This property is used in the reduction later.

\subsubsection{Colored graph to circuit}
Given a colored graph in which \textit{vertices are gates}, one can compose a circuit corresponding to the colored graph.

\begin{algorithm}[H]
	\small
	\caption{$\mathsf{ToCircuit}$}
	\begin{algorithmic}[1]
		\Require $\widetilde{G} = (V,E,c)$
		\Comment{$c:V\rightarrow C$, $C$ being a color set}
		\Ensure a sequence of gates $\v v$
		\algrule \vspace{2pt}
		\State Find a permutation $\rho : \{1,\ldots,|V|\} \rightarrow \{1,\ldots,|V|\}$ such that $c  (v_{\rho(i)}) \le c (v_{\rho(i+1)}) $
		\State \Return $(v_{\rho(1)} , v_{\rho(2)},\ldots,v_{\rho(|V|)})$
	\end{algorithmic}
	\label{alg:graph_to_circuit}
\end{algorithm}

\noindent Let $c_i$ be $c(v_i)$ for $i\in \{1,\ldots,|V| \}$.
What $\rho$ does is sorting $c_i$ such that smaller $c_i$ comes first.
We leave the sorting specification for $c_i = c_j$ to users, as whatever comes first does not affect the bound on the circuit depth (see below).
The output is self-explanatory; all gates colored with 1 are drawn first, followed by the gates colored with 2, and so on.
Therefore the depth is upper bounded by the number of colors used in $\widetilde{G}$.
The time complexity of the algorithm reads $O(|V| \log |V|)$.

The resulting circuit satisfies the following:
\begin{myRemark}
	The depth of the resulting circuit of Algorithm~\ref{alg:graph_to_circuit} is either (a) smaller than or (b) equal to the number of colors of the input graph.
	\label{remark:prop_Algo2}
\end{myRemark}

\noindent
Case\;(b) is straightforward, thus we leave a brief comment on the smaller case\;(a).
Given a colored graph, unconnected vertices correspond to parallelizable gates, but the colors on these vertices can be different.
Therefore, despite the different colors on the vertices, the corresponding gates from the output may happen to be in the same layer.
An example is given in Appendix\;\ref{sec:remark6}.

\subsection{Reduction}\label{sec:reduction}
Suppose we are given a commuting circuit $\v v$ and asked to reorder the gates such that its depth is minimized.
We construct a graph $G_{\cal V}$ corresponding to $\v v$ by Algorithm\;\ref{alg:construct_graph}.
The graph coloring problem is then solved by some solver, outputting a colored graph with the minimum number of colors used. 
The colored graph is then transformed back into a circuit by Algorithm\;\ref{alg:graph_to_circuit}.
The remaining task is to prove that the resulting circuit is the one with the minimum depth, completing the reduction from gate ordering problem to graph coloring problem.
More notions are introduced below for the task, but they are solely required for the proof.

Recall that we are given a circuit $\v v$, and $\cal V$ is the multiset of gates in $\v v$.
Define $A_{\cal V}$ as the set of all permutations of gates in $\cal V$.
Note that $\v v \in A_{\cal V}$.
Since we are optimizing a circuit by reordering the gates in $\v v$, the optimal circuit $\v v_{\rm opt}$ (which is not necessarily unique) also belongs to $A_{\cal V}$; $\v v_{\rm opt} \in A_{\cal V}$.

Recall that $G_{\cal V}$ is the graph drawn by Algorithm\;\ref{alg:construct_graph} for given $\v v$.
Let $B_{\cal V} = \mathbf{Col}(G_{\cal V})$, which is the set of all colored graphs of $G_{\cal V}$.

\begin{myClaim}
	Let $A_{\cal V}$ and $B_{\cal V}$ be as defined above.
	Define $m_{\rm depth}: A_{\cal V} \rightarrow \mathbb{Z}^{\ge 0}$ as a function computing the depth of a circuit $\v a \in A_{\cal V}$. Define $m_{\rm color}: B_{\cal V} \rightarrow \mathbb{Z}^{\ge 0}$ as a function counting the number of colors used in a colored graph $b \in B_{\cal V}$.
	We claim that there exist functions $f: B_{\cal V} \rightarrow A_{\cal V}$ and $g: A_{\cal V} \rightarrow B_{\cal V}$ such that,
	\begin{itemize}
		\item[$\bullet$] $\forall b \in B_{\cal V}$, $m_{\rm color}(b) \ge m_{\rm depth}(f(b))$,
		\item[$\bullet$] $\forall \v a \in A_{\cal V}$, $m_{\rm depth}(\v a) = m_{\rm color}(g(\v a))$.
	\end{itemize}
	\label{cla:functions}
\end{myClaim}

It will soon be shown that $f$ is Algorithm\;\ref{alg:graph_to_circuit}, and $g$ will be introduced later in this subsection.
Assuming the claim is correct, we have a straightforward proposition.

\begin{myProposition}
	If $m_{\rm color}(b) = \displaystyle\min_{b'\in B_{\cal V}} \Big( m_{\rm color}(b') \Big)$ holds, and so does $m_{\rm depth}(f(b)) = \displaystyle\min_{\v a' \in A_{\cal V}} \big( m_{\rm depth}(\v a') \big)$.
	\label{prop:reduction}
\end{myProposition}
\begin{proof}
	We use the proof by contradiction.
	Let $b \in B_{\cal V}$ be the colored graph such that
	\begin{align}\label{eq:reduction-1}
	m_{\rm color}(b) = \min_{b'\in B_{\cal V}} \big(m_{\rm color}(b') \big) \enspace.
	\end{align}
	Now suppose that $m_{\rm depth}(f(b)) > \min_{\v a' \in A_{\cal V}} \big( m_{\rm depth}(\v a') \big)$, then there exists $\v a \in A_{\cal V}$ such that $m_{\rm depth}(\v a) < m_{\rm depth}(f(b))$, $\v a \neq f(b)$.
	
	Due to the properties of $f$ and $g$, we have $m_{\rm depth}(\v a) = m_{\rm color}(g(\v a))$ and $m_{\rm depth}(f(b)) \le m_{\rm color}(b)$.
	In other words, there exists $g(\v a) \in B_{\cal V}$ satisfying
	\begin{align*}
	m_{\rm color}(g(\v a)) = m_{\rm depth}(\v a) < m_{\rm depth}(f(b)) \le m_{\rm color}(b) \enspace,
	\end{align*}
	which contradicts to Eq.\;(\ref{eq:reduction-1}).
	%
\end{proof}


Proposition\;\ref{prop:reduction} tells us that if a colored graph $b$ is a graph achieving the chromatic number, then the circuit $f(b)$ is guaranteed to be a circuit with the minimum depth in $A_{\cal V}$.

Now we prove the claim.
It has already been shown that Algorithm\;\ref{alg:graph_to_circuit} satisfies the property $\forall b \in B_{\cal V}:$ $m_{\rm color}(b) \ge m_{\rm depth}(\mathsf{ToCircuit}(b))$ as in Remark\;\ref{remark:prop_Algo2}.
The function $g$ in Claim\;\ref{cla:functions} is given below.
Let us handle a coloring function $c$ as a table and fill its entry $c_i$ one by one so that at the end of the algorithm we have $c(v_i) = c_i$ for all $i\in \{1,\ldots,|V|\}$.

%
\begin{algorithm}[H]
	\small
	\caption{$\mathsf{ToColoredGraph}$}
	\begin{algorithmic}[1]
		\Require $\v v = (v_{1}, \ldots, v_{n}) \in A_{\cal V}$
		\Ensure $\widetilde{G}_{\cal V} = (V,E,c) \in B_{\cal V}$
		\algrule
		\State $(V, E) \gets \mathsf{ConstructGraph}(\v v)$
		\Comment {$V = \{v_{1}, \ldots, v_{n}\}$}
		\State $i \gets 1$, $c_1 \gets 1$, $d \gets 1$
		\Comment {$d$ for depth}
		\While {$i < n$}
		\State $j \gets i+1$
		\While {$j \le n$}
		\If {$E \cap \{ e_{kl} : k,l \in \{i, \ldots, j\}\} = \varnothing$}
		\State $c_{j} \gets c_i$
		\State $j \gets j+1$
		\Else
		\State $c_{j} \gets c_i + 1$
		\State $d \gets d + 1$
		\State \textbf{break}
		\EndIf
		\EndWhile
		\State $i \gets j$
		\EndWhile
		\State \Return $(V,E,c)$
		\Comment {$c(v_i) = c_{i}$}
	\end{algorithmic}
	\label{alg:circuit_to_graph}
\end{algorithm}
\noindent
%
\noindent Let $\v v$ be an input to $\mathsf{ToColoredGraph}$. 
Recall that the vertices are gates. 
Algorithm~\ref{alg:circuit_to_graph} sequentially identifies the parallelizable gates (vertices), and those vertices are identically colored in Line 7.
Whenever a new color is introduced in Line 10, $d$ is also increased by 1.
Since $d$ is increased only when $v_j$ is non-parallelizable with any of $v_i,\ldots,v_{j-1}$ (deducible from Line 6), the final value of $d$ at the end is equal to the depth of the circuit.
Therefore, we see that $\forall \v a \in A_{\cal V}$, $m_{\rm depth}(\v a) = m_{\rm color}(\mathsf{ToColoredGraph}(\v a))$.
Time complexity of Algorithm\;\ref{alg:circuit_to_graph} is $O(n^{2})$. 

We have shown that Claim\;\ref{cla:functions} does hold, thereby completing the reduction from gate ordering to graph coloring.
A reduction for the opposite direction, from graph coloring to gate ordering, is less important as the main purpose of this work is to suggest a strategy for circuit optimization.
Therefore we leave the opposite reduction in Appendix~\ref{sec:reduction-other-way}.

\subsection{Circuit optimization strategy}
The fact that gate ordering is reduced to graph coloring results in a way to optimize the depth of a commuting circuit.
A simple strategy is to utilize any (approximate) algorithm for solving graph coloring.

\begin{enumerate}
	\item Given a circuit, construct the corresponding graph by using Algorithm\;\ref{alg:construct_graph}.
	\item Solve graph coloring by using known solvers. The output is a colored graph.
	\item Compose a circuit corresponding to the colored graph by using Algorithm\;\ref{alg:graph_to_circuit}.
\end{enumerate}

It can be deduced from inspecting the reduction steps that an approximate solution of graph coloring leads to at least as good approximations to the gate ordering problem.
Let $\v v$ be a commuting circuit, $\v v_{\rm opt}$ be the optimal circuit of $\v v$, $b$ be an approximate solution for the graph coloring problem of $\mathsf{ConstructGraph}(\v v)$, and let $h' = m_{\rm color}(b)$.
Since $b$ is not optimal, we have $\Delta := h' - h \ge 0$, where $h$ is the chromatic number of the graph.
Due to Remark\;\ref{remark:prop_Algo2}, we see that $m_{\rm depth}(\mathsf{ToCircuit}(b)) - m_{\rm depth}(\v v_{\rm opt}) \le \Delta$, meaning that any approximate solution $b$ of the graph coloring leads to a circuit with smaller or equal gap to the minimum depth.
Note however that the above argument holds only for approximate solutions. 
If the coloring is optimal, then the depth of the corresponding circuit is equal to the number of colors.

Since graph theory has been studied extensively over history, one may expect the strategy to be effective on some occasions.
Computational complexity and the depth of the resulting circuit are determined by the algorithm for graph coloring.





\section{Validation}\label{sec:algorithms}
In this section, the validity of the reduction strategy is put to the test by implementing Algorithm\;\ref{alg:circuit-opt}.
Any coloring algorithm can be used, and we choose two well-known algorithms (solvers) and evaluate their performance.

\begin{algorithm}[H]
	\small
	\caption{$\mathsf{OptimizeCircuit}$}
	\begin{algorithmic}[1]
		\Require $\v v$
		\Comment{input circuit}
		\Ensure $\v v'$
		\Comment{output circuit}
		\algrule \vspace{2pt}
		\State $G \gets \mathsf{ConstructGraph}(\v v)$
		\State $\widetilde{G} \gets \mathsf{ColoringSolver}(G)$
		\State $\v v' \gets \mathsf{ToCircuit}(\widetilde{G})$
		\State \Return $\v v'$
	\end{algorithmic}
	\label{alg:circuit-opt}
\end{algorithm}

\subsection{Coloring solvers}
\subsubsection{DSatur}
DSatur is a saturation-degree-based greedy algorithm\;\cite{ColDSatur}.
Saturation degree of a vertex is the number of different colors used by its neighbors.
Roughly described, at each step of the algorithm, an uncolored vertex with the highest degree is selected and assigned a color such that the use of a new color is suppressed if possible.

The time complexity is easily estimated.
Let the number of vertices of an input graph $G$ be $n$.
At each step, for a selected vertex, the algorithm inspects only its neighbors and then colors it.
It repeats for all $n$ vertices, thus the running time is bounded by $O(n^2)$.

About the number of colors of the output, let $\chi_G$ be the chromatic number of $G$.
The algorithm is heuristic, and there is no general theoretical bound that characterizes how close DSatur's output is to $\chi_G$.
However, the empirical performance of DSatur has been studied extensively, and the benchmark results can be found in various literature\;\cite{DSaturBenchmark1,DSaturBenchmark2}.

\subsubsection{Backtracking}
Backtracking is an algorithm for $k$-coloring, which asks if a given graph is colorable with at most $k$ different colors\;\cite{backtracting}.
Basically, Backtracking performs a depth-first search over color assignments.
It colors vertices one by one while respecting constraints. 
When a vertex cannot be assigned a color anymore without violating the constraints, the algorithm backtracks to change earlier color choices, effectively pruning that branch of the search space.
The worst-case time complexity is known to be $O(k^n)$ for $k$-coloring and it can be used to find the chromatic number by trying different $k$ at most $n$ times.

\subsection{Lower bound on the depth of a commuting circuit}
In Section\;\ref{sec:main}, we have not specified conditions for operations to be simultaneously executable since the parallelizability depends on the underlying hardware.
To analyze the algorithm however, the condition has to be fixed.
We begin with the following assumption that two gates sharing no common qubit can be executed simultaneously:

\begin{myAssumption}\label{assump:parallelizability}
	Let $v, w$ be unitary operations and $Q_v$, $Q_w$ be sets of qubits $v$ and $w$ operate on, respectively.
	The operations $v, w$ are simultaneously executable if and only if $Q_v \bigcap Q_w = \varnothing$.
\end{myAssumption}

When a commuting circuit is generated randomly (each gate is chosen randomly), the corresponding graph by Algorithm\;\ref{alg:construct_graph} becomes a \textit{random intersection graph}\;\cite{RIG99}.
We skip the details on the model, but take the bound on the chromatic number.
Let $m$ be the number of qubits of a random commuting circuit and $G$ be the corresponding graph, then
\begin{align}\label{eq:RIG-bound}
\chi(G) \ge \max_{1\le q\le m} |L_q|\enspace,
\end{align}
where $\chi(G)$ is the chromatic number of $G$, and $L_q$ is the set of gates that act on the $q$-th qubit.
For example for a circuit $\v v_2 = (C\!X_{134}, C\!X_{235}, C\!X_{167}, C\!X_{268})$, we have $L_1=\{C\!X_{134}, C\!X_{167}\}$ and $L_2=\{C\!X_{235}, C\!X_{268}\}$.
Briefly explained, the largest set $L_q$ naturally forms the maximum clique\;\cite{bondy2008graph} whose size provides the non-tight lower bound on the chromatic number.

It is worth noting that for certain parameter regimes- the number of gates, the number of qubits, the sizes of gates- the bound is nearly tight and polynomial time algorithms can meet the lower bound\;\cite{BTU09,KR15,Ryb17}.
However, delving into the optimal solvability of commuting circuits is beyond the scope of this work.



\subsection{Experiments}
\begin{figure}[htbp]
	\centering
	\includegraphics[width=\textwidth]{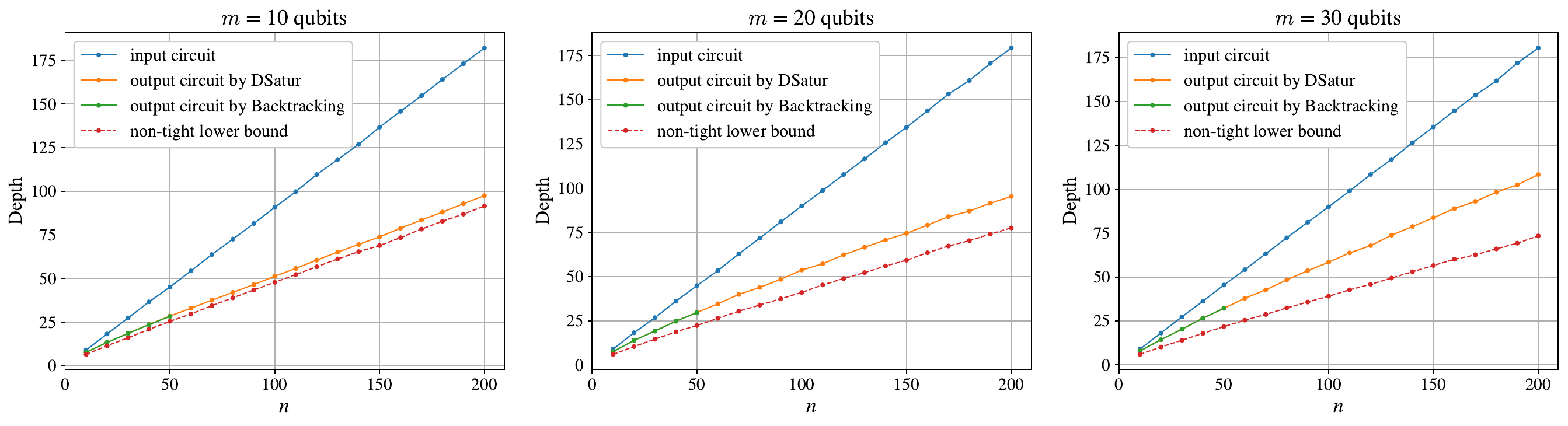}
	\caption{Optimization by coloring for randomly generated commuting circuits with $m \in \{10,20,30\}$ qubits and $n \in \{10,20,...,200\}$ gates. Blue, orange, green, and red markers denote, respectively, the averaged depths of the input circuit, the output circuit produced by DSatur, the output circuit produced by Backtracking, and the non-tight lower bound by Eq.\;(\ref{eq:RIG-bound}).}
	\label{fig:experiemnt}
\end{figure}

We generate commuting circuits with $m \in \{10,20,30\}$ qubits and $n \in \{10,20,...,200\}$ multi-controlled $Z$ gates.
When generating a gate, we randomly choose at least three to at most $\lfloor m/2 \rfloor$ qubits the gate will act on.
The generated circuit is then fed to Algorithm\;\ref{alg:circuit-opt} as an input.
For each $m$ and $n$, 100 circuits are generated and optimized by the algorithm.
The results are shown in Figure\;\ref{fig:experiemnt}.

In Figure\;\ref{fig:experiemnt}, each data point represents an average over 100 independent trials.
The blue marker indicates the depth of the input circuits, the orange (green) marker represents the depth of the output circuit obtained using the DSatur (Backtracking) solver, and the red marker denotes the non-tight lower bound given by Eq.\;(\ref{eq:RIG-bound}).
Since the running time of the exact solver is exponential in the number of gates, the green markers are present up to $n=50$.
For all $(n,m)$ values where green markers are present, the orange markers lie unnoticeably slightly above or exactly on the green ones. 
How much more the two solvers deviate for larger circuit sizes is open for further experiment.
Each set of circuit parameters ($n$, $m$, and gate sizes) may correspond to a distinct hardness regime for graph coloring in random intersection graphs\;\cite{Ryb17}.
Based on the observations, we hypothesize that gate ordering is easier for small numbers of qubits, as the polynomial time heuristic algorithm produced circuits that were nearly optimal.
For larger numbers of qubits however, we cannot immediately draw the conclusion that the problems are harder since the lower bound is non-tight.

In addition, we conducted an experiment on the coloring algorithms within the Qiskit framework.
Direct comparison is limited because Qiskit transpiler does not work as a gate ordering solver.
Instead, we let Qiskit to decompose a commuting circuit, with and without the coloring algorithms applied, into Clifford+$T$ circuit and then let it optimize to reduce the number of $T$ gates.
Details are given in Appendix\;\ref{sec:Qiskit}.



\section{Applications}\label{sec:applications}

It is clear that the proposed optimization strategy can be utilized to reduce the depth of instantaneous quantum polynomial or commuting local Hamiltonian circuits since these circuits consist only of commuting gates.
In addition, commuting circuits are commonly found in quantum algorithms that require superposed queries.
Let $f$ be a function and consider its superposed query,
\begin{align*}
\sum_x \ket{x} \ket{0} \mapsto \sum_x \ket{x} \ket{f(x)} \enspace.
\end{align*}
Implementation of such query often involves commuting gates, as the qubits in the first and the second registers are mainly used for controls and targets, respectively.
Two such applications are examined in this section including finite field multiplications and integer additions based on quantum Fourier transform.


\subsection{Finite field multiplication}\label{sec:application-1}
A quantum circuit for multiplication in $\mathbb{F}_{p^n}$ for prime $p$ and $n\in \mathbb{Z}^+$ is an important component in various quantum algorithms.
Researchers have developed various designs for finite field multiplications\;\cite{MulMaslov,mulGNB,mulghost,RS14,mulKepley,mulSpaceEfficient,mulCF}, and here we focus on the one proposed by Maslov et al., which involves the minimum number of qubits\;\cite{MulMaslov}.
The result for $p=2$ is summarized as follows:

\begin{figure}[htbp]
	\centering
	\includegraphics[width=0.54\textwidth]{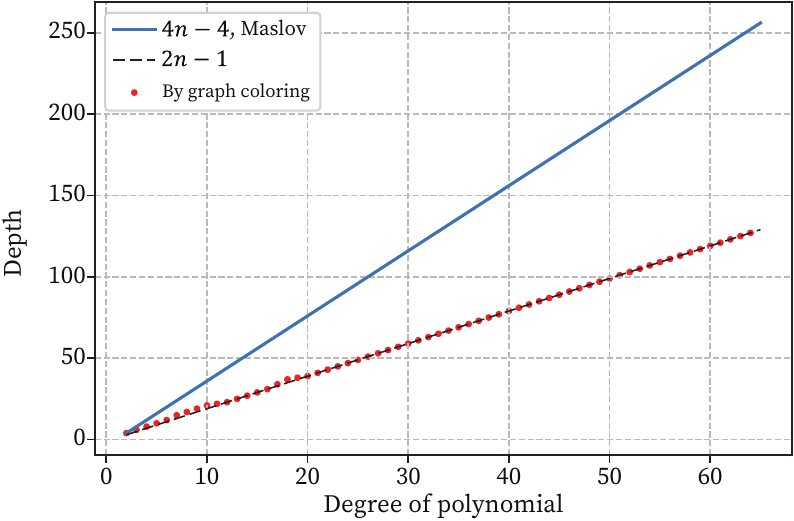}
	\caption{Depth optimization of multiplication in $\mathbb{F}_{2^n}$ by solving graph coloring. Red markers are obtained by using DSatur algorithm for graph coloring, black dashed line is the lower bound, and blue solid line is Maslov et al's original design. } 
	\label{fig:mult-coloring}
\end{figure}

\begin{myRemark}
	Using graph coloring, Toffoli-depth of the multiplication circuit is reduced to as small as $2n-1$, from $4n-4$.
\end{myRemark}
%



Maslov et al. originally proposed a circuit for binary field multiplication, but it can easily be extended to any $p>2$. 
The binary field multiplication circuit takes as inputs three quantum registers holding $a\in \mathbb{F}_{2^n}$, $b\in \mathbb{F}_{2^n}$, and $0$, respectively on $n$ qubits each, and outputs $a,b,a\cdot b$.
It requires $n^2$ Toffoli gates and $O(n^2)$ CNOT gates, with $4n-4$ Toffoli-depth.
An important feature of the design is that no extra qubit is involved throughout the operation other than the initial $3n$ qubits.

\begin{figure}[htbp]
	\centering
	\includegraphics[width=0.6\textwidth]{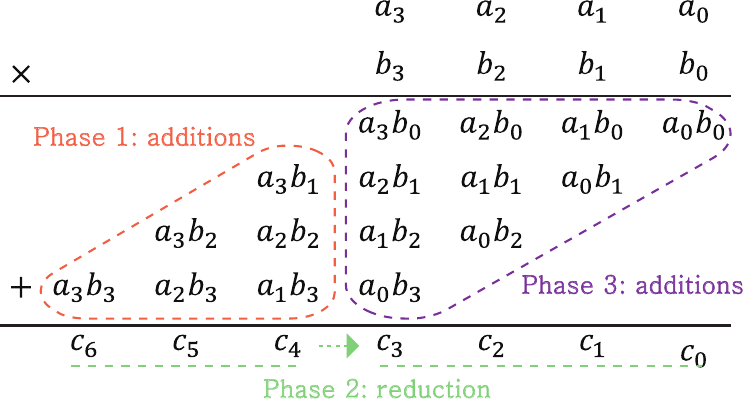}
	\caption{Illustration of the multiplication. The terms in each triangle are to be summed where the summing order does not affect the result.}
	\label{fig:mult-table}
\end{figure}

The design approach is reviewed first, and then we point out where the graph coloring can be exploited.
For our purpose, it suffices to review a simple example given by\;\cite[Section 3]{MulMaslov}.
Consider multiplying $a \cdot b$ in $\mathbb{F}_{2^4}$ with the irreducible reducing polynomial $P(x) = x^4 + x + 1$.
Let $\v a$ (and also $\v b$) be a vector of coefficients of the polynomial; $\v a = (a_3 , a_2 ,a_1 ,a_0)$, $\v b = (b_3 , b_2 ,b_1 ,b_0)$.
Define
\begin{align*}
\v d
=
\left[\begin{array}{c}
a_0 b_0 \\
a_1 b_0 + a_0 b_1 \\
a_2 b_0 + a_1 b_1 + a_0 b_2 \\
a_3 b_0 + a_2 b_1 + a_1 b_2 + a_0 b_3
\end{array}
\right],
~
\v e
=
\left[\begin{array}{c}
a_3 b_1 + a_2 b_2 + a_1 b_3 \\
a_3 b_2 + a_2 b_3 \\
a_3 b_3
\end{array}
\right],
~
Q
=
\left[\begin{array}{ccc}
1 & 0 & 0 \\
1 & 1 & 0 \\
0 & 1 & 1 \\
0 & 0 & 1 \\
\end{array}
\right] \enspace,
\end{align*}
and then one can verify that $\v a \cdot \v b = \v d + Q \v e$ (see,\;\cite{MulMaslov} for details).
Inspecting the equality, quantum multiplication can be divided into three phases.
Phase\;1 is computing $\v e$ on the empty register, Phase\;2 is multiplying (reducing) the result by $Q$, and Phase\;3 is adding the result by $\v d$.
Writing down it as a unitary transformation,
\begin{align*}
\ket{\v a}\ket{\v b}\ket{0}
\stackrel{\text{\tiny Phase 1}}{\longmapsto}
\ket{\v a}\ket{\v b}\ket{\v e}
\stackrel{\text{\tiny Phase 2}}{\longmapsto}
\ket{\v a}\ket{\v b}\ket{Q \v e}
\stackrel{\text{\tiny Phase 3}}{\longmapsto}
\ket{\v a}\ket{\v b}\ket{\v d + Q \v e} \enspace.
\end{align*}
Figure\;\ref{fig:mult-table} illustrates the procedures.
Focusing on the Toffol-depth, in $\mathbb{F}_{2^n}$, the authors estimated the depth to be $2n -3$, $0$, $2n -1$ in Phase 1, Phase 2, and Phase 3, respectively, in total $4n-4$.

Now notice that all the additions occurring in Phase\;1 are commuting.
Similarly, all the additions in Phase\;3 are also commuting.
The graph coloring strategy is applied to each Phase\;1 and Phase\;3 in $\mathbb{F}_{2^n}$ for $n \le 512$, adopting the solver DSatur which is a polynomial time heuristic algorithm for graph coloring\;\cite{ColDSatur}.
Total Toffoli-depth is the sum of the coloring results in two phases.
Figure\;\ref{fig:mult-coloring} summarizes the result.
(Plots for $n>64$ are not shown in the figure, but all of them do not deviate at all from $2n-1$.)
For $n\ge 6$, a lower bound on the Toffoli-depth of $2n-1$ can be deduced from an elementary argument,\footnote{It is the sum of the lengths of the bases of two triangles in Figure\;\ref{fig:mult-table}. For $n< 6$, the lower bound is larger than $2n-1$ since a right triangle of base length $\le 4$ (which appears for $n\le 5$) cannot be fully parallelized. } and DSatur has found circuits meeting the lower bound for most $n$ tested.\footnote{For some small $n\; (<20)$ the corresponding graphs are rather denser in edge density, and the adopted DSatur code~\cite{pyqubo} has failed to meet the bound.}

The above argument also holds in non-binary fields, $\mathbb{F}_{p^n}$ with prime $p>2$.
The difference is that the arithmetics of coefficients take place in the base field.
Such arithmetics themselves may not consist entirely of commuting gates, but the addition operations are still commuting, allowing the optimization strategy to be applicable.


\subsection{Depth-qubit tradeoffs for QFT-based addition}\label{sec:application-2}
\begin{figure}[htbp]
	\centering
	\includegraphics[width=0.95\textwidth]{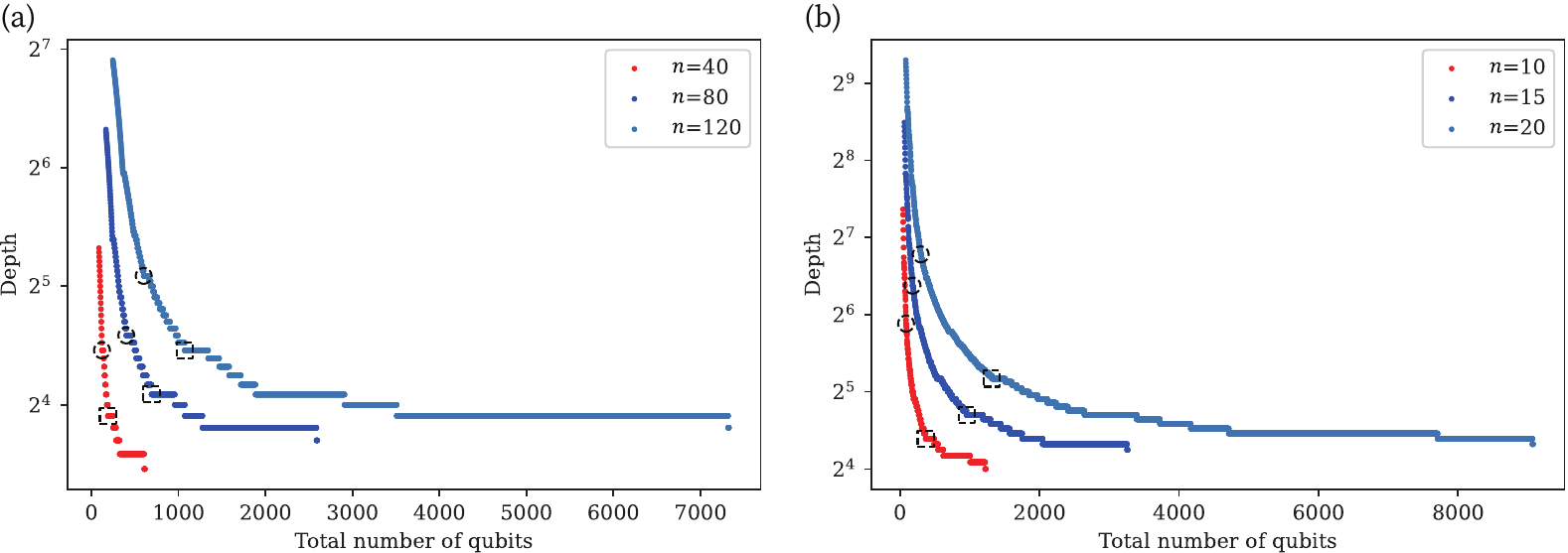}
	\caption{Depth-qubit tradeoffs for QFT-based addition circuits (a) $\ket{a}\ket{b} \mapsto \ket{a}\ket{b+a}$ and (b) $\ket{a}\ket{b}\ket{c}\mapsto \ket{a}\ket{b}\ket{c+ab}$, where $a$ and $b$ are $n$-bit numbers and $c$ is set to zero. The circled and squared markers are the most efficient circuit designs with respect to the cost metrics in Eq.\;(\ref{eq:metric}), respectively (see, Table\;\ref{tab:cost-comparison}).}
	\label{fig:QFT-coloring}
\end{figure}

The graph coloring strategy can provide depth-qubit tradeoff options.
As an example, we look for depth-qubit tradeoffs for QFT-based addition developed by Draper\;\cite{draper} and its variant\;\cite{RG17}.
Having the original design as one extreme end (qubit saving), the goal is to find circuits with various numbers of qubits and depths.
The result is summarized in Figure\;\ref{fig:QFT-coloring}.



\begin{figure}[htbp]
	\centering
	\includegraphics[width=0.9\textwidth]{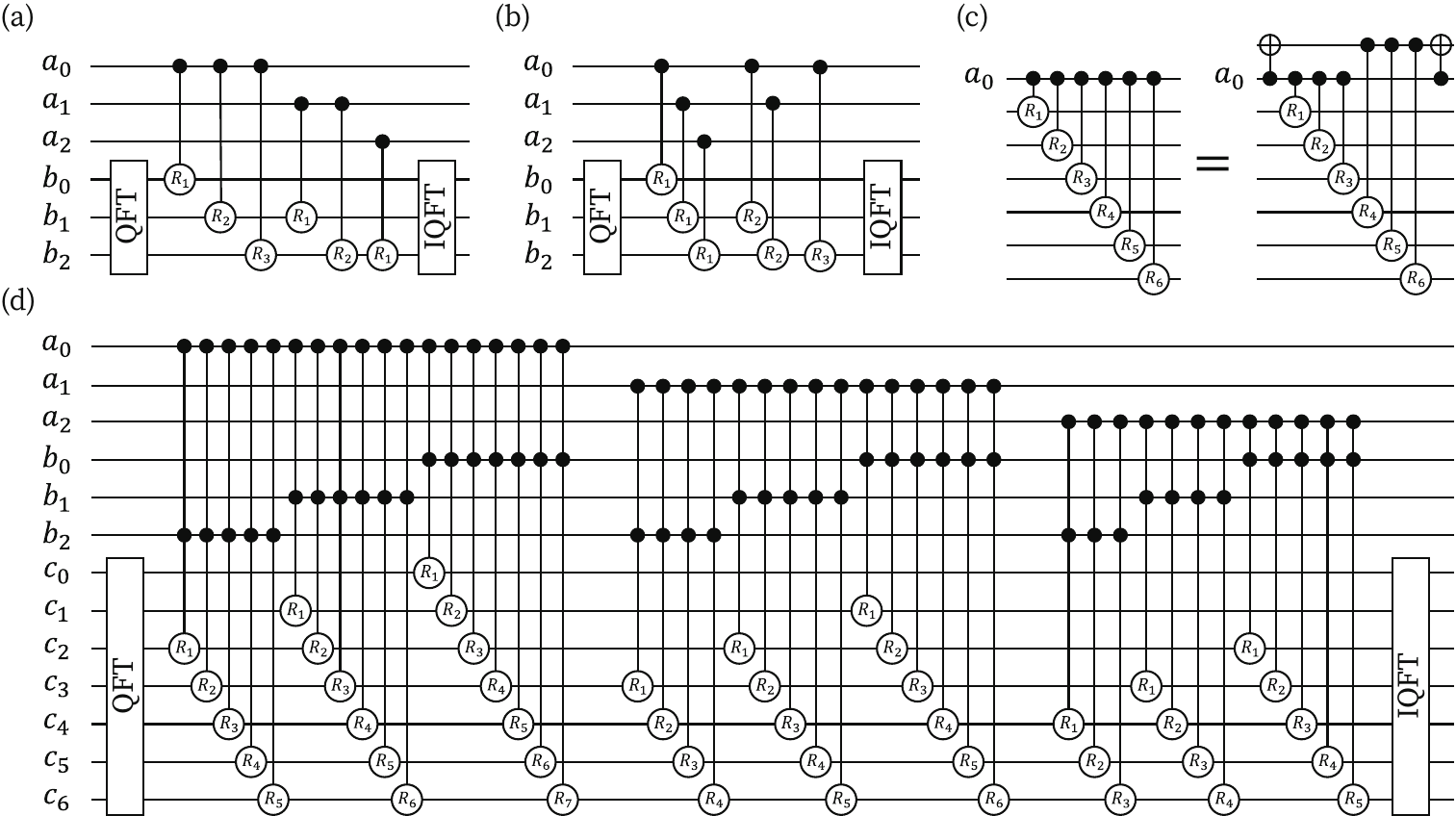}
	\caption{QFT-based addition $\ket{a}\ket{b} \mapsto \ket{a}\ket{b+a}$ is shown in (a) and its (trivially) optimized circuit is given in (b). Reducing the depth by extra work qubit is illustrated in (c).
		A circuit for $\ket{a}\ket{b}\ket{c}\mapsto \ket{a}\ket{b}\ket{c+ab}$ is given in (d).}
	\label{fig:QFT-add-circuits}
\end{figure}

Briefly reviewed, Draper's design for the addition works as follows.
Consider adding two numbers $a,b \in \mathbb{Z}$, $\ket{a}\ket{b} \mapsto \ket{a}\ket{b+a}$.
To this end, we first carry out the QFT on the second register, $\mathsf{QFT} \ket{b} = \ket{\phi (b)}$.
Having qubits in $\ket{a}$ register as controls, conditional rotation operations are applied such that $\ket{a}\ket{\phi(b)} \mapsto \ket{a}\ket{\phi(b+a)}$.
Finally, the inverse QFT transforms the second register back into the desired result, $\mathsf{QFT^{-1}} \ket{\phi(b+a)} = \ket{b+a}$.
Figure\;\ref{fig:QFT-add-circuits}\,(a) describes an example.

Notice that excluding the two QFTs, the conditional rotation gates are all commuting.
It is thus possible to optimize the circuit by solving the graph coloring, but the optimal depth (excluding two QFTs) is trivially inferred as $n = \max (\lceil \log_2 a \rceil, \lceil \log_2 b \rceil )$ by observation.
Instead of optimizing the circuit itself, one may consider further reducing the depth by introducing extra qubits.
The basic idea is that if a specific qubit is referred to multiple times as controls or targets contributing a substantial amount to the overall depth, and then sharing its role (as control or target) with $x$ more qubits reduce the relevant depth to around $1/(x+1)$.
For example, introducing one initialized qubit in Figure\;\ref{fig:QFT-add-circuits}\,(c) reduces the depth by half (except the two CNOT gates at both ends) by \textit{assisting} $a_0$.
We call a circuit optimization by the above method parallelization by extra qubits.

An algorithm for the parallelization by extra qubits can be devised by using graph coloring.
A heuristic algorithm is developed in which the central idea is noted below.
(We do not elaborate on the algorithm as the algorithm itself is not the key finding.)

\begin{figure}[htbp]
	\centering
	\includegraphics[width=0.9\textwidth]{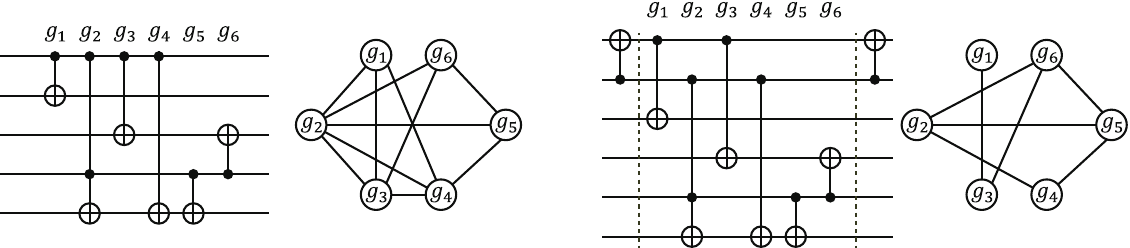}
	\caption{An example of the parallelization by extra qubits. Introducing one extra qubit gives rise to altering the graph such that it becomes sparser (in edge density). }
	\label{fig:tradeoffs-idea}
\end{figure}

Consider a commuting circuit and its corresponding graph $G = (V,E)$.
Since gates are vertices, introducing extra qubits leaves $V$ unchanged, but depending on which gates are lifted, $E$ changes such that the number of connected edges decreases.
Figure\;\ref{fig:tradeoffs-idea} shows how the graph can be altered by one extra qubit.
The key idea of the algorithm is that as we cut out the edges from the graph, the chromatic number can only get smaller.
For example, let $G(n,p)$ be a random graph, where $n$ is the number of vertices and $p$ is the probability that an edge exists between two vertices.
Bollob\'{a}s has shown that the chromatic number of a random graph $\chi(G(n,p))$ asymptotically reads\;\cite{Bol98},
\begin{align*}
\chi(G(n,p)) \sim \frac{n}{2 \log_{1/(1-p)} n} (1+o(1)) \enspace.
\end{align*}
In other words, the chromatic number of a random graph gets smaller as $p$ approaches 0.
Therefore, we use the extra qubits in a way that the altered graph becomes as sparse in edge density as possible.\footnote{Beware that not every commuting circuit corresponds to a random graph or a random intersection graph.}
By using the algorithm, we have parallelized the QFT-based additions $\ket{a}\ket{b} \mapsto \ket{a}\ket{b+a}$ and $\ket{a}\ket{b}\ket{c} \mapsto \ket{a}\ket{b}\ket{c+ab}$.
For simplicity, we set $\lceil \log_2 a \rceil = \lceil \log_2 b \rceil = n$ and $c=0$.
The results are summarized in Figure\;\ref{fig:QFT-coloring}.
(Up to roughly 10,000 qubits are considered, far exceeding what current hardware can support at the time of writing. One should take it as a theoretical calculation.)

\begin{table}[H]
	\centering
	\caption{Cost ratio of $C_{\rm space}$ and $C_{\rm time}$ to the most cost-efficient circuit obtained by the algorithm, where $C_{\rm space}$ and $C_{\rm time}$ are the most space- and time-saving circuits for QFT-based additions.
		Two different metrics $S$ and $T$ in Eq.\;(\ref{eq:metric}) are examined.
		The greater the value is than 1, the more inefficient it is.
		In Figure\;\ref{fig:QFT-coloring}, in each tradeoffs curve, the circuits for $\min_{C}S(C)$ and $\min_{C}T(C)$ are highlighted by dashed circles and squares, respectively.}
	{\renewcommand{\arraystretch}{1.2}
		\begin{tabular}{>{\centering}p{2.9cm} |
				>{\centering}p{1.4cm} >{\centering}p{1.4cm} >{\centering}p{1.4cm} | >{\centering}p{1.4cm} >{\centering}p{1.4cm} >{\centering}p{1.4cm}
			}
			& \multicolumn{3}{c|}{Figure\;\ref{fig:QFT-coloring}\,(a)} & \multicolumn{3}{c}{Figure\;\ref{fig:QFT-coloring}\,(b)} \\[2pt] \cline{2-7}
			& $n=40$ & $n=80$ & $n=120$ & $n=10$ & $n=15$ & $n=20$ \tabularnewline \hline
			{\small $\displaystyle S(C_{\text{space}}) / \min_{C}S(C) $}& 1.21 & 1.33 & 1.41 & 1.33 & 1.48 & 1.57 \tabularnewline
			{\small $\displaystyle S(C_{\text{time}}) / \min_{C}S(C) $}& 2.53 & 3.50 & 5.02 & 3.85 & 4.16 & 5.57 \tabularnewline \hline
			{\small $\displaystyle T(C_{\text{space}}) / \min_{C}T(C) $}& 3.03 & 5.14 & 6.66 & 6.90 & 12.27 & 18.84 \tabularnewline
			{\small $\displaystyle T(C_{\text{time}}) / \min_{C}T(C) $}& 1.73 & 2.19 & 2.77 & 1.93 & 1.82 & 2.12 \tabularnewline \hline
		\end{tabular}
	}
	\label{tab:cost-comparison}
\end{table}

The tradeoff designs allow one to choose more economical options when the total cost of running a circuit is measured in terms of the depth (time cost) and the number of qubits (space cost).
For example consider two cost functions $S$ and $T$,
\begin{align}\label{eq:metric}
S(C) = \mathsf{depth}(C) \cdot \mathsf{qubit}(C),
~~~~
T(C) = (\mathsf{depth}(C))^2 \cdot \mathsf{qubit}(C) \enspace,
\end{align}
where $C$ is a circuit, $\mathsf{depth}(C)$ is the depth of $C$, and $\mathsf{qubit}(C)$ is the number of qubits required to run $C$.
The function $S$ puts equal weights on time and space, whereas $T$ values time more than space as it requires quadratically more space to reduce time as in the quantum search algorithm\;\cite{zalka}.
Based on the two metrics $S$ and $T$, the cost of two extreme designs (the most space- and time-saving ones) for QFT-based additions is compared with the most cost-efficient circuit obtained by the algorithm.
Table\;\ref{tab:cost-comparison} summarizes the results.
For example, in QFT-addition with $n=120$, $S(C_{\text{space}}) / \min_{C}S(C) = 1.41$ means the most space-saving circuit is 1.41 times more expensive than the optimized one found by the algorithm when considering the cost metric $S$.
The circuit with the lowest cost is highlighted in Figure\;\ref{fig:QFT-coloring}.




\section{Conclusion and future directions}\label{sec:summary}
In this work, optimizing the depth of a commuting circuit is reduced to solving graph coloring problem, leading to a strategy for circuit optimization.
It has been shown that the strategy can be applied to improve the known circuits.

One limitation of the reduction strategy is that it only works on commuting circuits.
Two workarounds can be further investigated.
One direction is to transform a noncommuting circuit into one such that a portion of commuting parts is large.
This is closely related with the circuit synthesis where the behavior of an input circuit should be maintained in the output\;\cite{nam18,MQT}.
The other way is to divide a noncommuting circuit into many commuting parts and then applying the method where available.
For example, LU decomposition is employed in optimizing linear circuits\;\cite{PLU_SODA,MZ22}, where we see a potential application of the coloring method.

We want to point out that the strategy developed in this work is capable of finding the optimal depth of a commuting circuit corresponding to a perfect graph\;\cite{bondy2008graph}, since there exists a polynomial time algorithm for graph coloring of a perfect graph\;\cite{PolyTimePerfectColoring}.
The next question is then whether it is possible to find such a circuit that has practical uses while its optimal depth is not yet known.
We have not formally proved yet, but considering that the perfect graphs do not have odd ring of length $\ge 5$, a circuit consisting of commuting CNOT gates seems to correspond to a perfect graph.\footnote{On the other hand, CZ-only circuits generally do not correspond to perfect graphs.}
However, we were not able to take advantage of such a fact for a useful application.

An anonymous reviewer raised a question whether the reduction idea can be generalized to account for hardware restrictions such as qubit connectivity and gate set.
If hardware restrictions are to be considered, an originally commuting circuit will naturally become noncommuting one.
We have tried to generalize the reduction idea so that it works on such restrictions, but came to the conclusion that it is highly nontrivial.
There is a possibility that vertex coloring of a certain acyclic directed graph may properly reflect hardware restrictions, but we were not able to think of a clear and efficient way.

An anonymous reviewer raised another question if the vertex coloring approach is advantageous over the edge coloring of a hypergraph.
In our estimation, it seems possible to find similar reduction steps by using the edge coloring in an appropriate hypergraph.
To compare the better solvability of the edge coloring and the vertex coloring is an interesting but nontrivial task, since the target graphs as well as the known solvers are different.
For example, the reduced hypergraph for edge coloring would have the same number of vertices and edges as the number of qubits and gates, respectively, whereas the reduced graph for vertex coloring has the same number of vertices as the number of gates but its edges are not directly relevant to the number of qubits.
Resolving the above issues may be worth independent work.

\section*{Acknowledgement}
We are grateful to Raghav Govind Jha for introducing the partitioning of Pauli strings and the relevant papers.

\let\v=\oldv
\bibliographystyle{quantum}
\bibliography{reference}

\begin{thebibliography}{10}

\bibitem{IQP}
Dan Shepherd and Michael~J. Bremner.
\newblock ``{Temporally unstructured quantum computation}''.
\newblock \href{https://dx.doi.org/10.1098/rspa.2008.0443}{Proceedings of the
  Royal Society A: Mathematical, Physical and Engineering Sciences {\bf 465},
  1413--1439}~(2009).

\bibitem{IQP2}
Michael~J. Bremner, Ashley Montanaro, and Dan~J. Shepherd.
\newblock ``{Achieving quantum supremacy with sparse and noisy commuting
  quantum computations}''.
\newblock \href{https://dx.doi.org/10.22331/q-2017-04-25-8}{{Quantum} {\bf 1},
  8}~(2017).

\bibitem{CLH}
Sergey Bravyi and Mikhail Vyalyi.
\newblock ``{Commutative version of the local Hamiltonian problem and common
  eigenspace problem}''.
\newblock \href{https://dx.doi.org/10.26421/qic5.3-2}{Quantum Info. Comput.
  {\bf 5}, 187–215}~(2005).

\bibitem{toric}
A.Yu. Kitaev.
\newblock ``{Fault-tolerant quantum computation by anyons}''.
\newblock \href{https://dx.doi.org/10.1016/S0003-4916(02)00018-0}{Annals of
  Physics {\bf 303}, 2--30}~(2003).

\bibitem{MulMaslov}
Dmitri Maslov, Jimson Mathew, Donny Cheung, and Dhiraj~K. Pradhan.
\newblock ``{An $O(m^2)$-depth quantum algorithm for the elliptic curve
  discrete logarithm problem over $GF(2^m)$}''.
\newblock \href{https://dx.doi.org/10.26421/qic9.7-8-4}{Quantum Info. Comput.
  {\bf 9}, 610–621}~(2009).

\bibitem{draper}
Thomas~G. Draper.
\newblock ``{Addition on a quantum computer}''~(2000).
\newblock
  \href{http://arxiv.org/abs/quant-ph/0008033}{arXiv:quant-ph/0008033}.

\bibitem{SM13}
Mehdi Saeedi and Igor~L. Markov.
\newblock ``{Synthesis and optimization of reversible circuits—a survey}''.
\newblock \href{https://dx.doi.org/10.1145/2431211.2431220}{ACM Comput.
  Surv.{\bf 45}}~(2013).

\bibitem{nam18}
Yunseong Nam, Neil~J Ross, Yuan Su, Andrew~M Childs, and Dmitri Maslov.
\newblock ``{Automated optimization of large quantum circuits with continuous
  parameters}''.
\newblock \href{https://dx.doi.org/10.1038/s41534-018-0072-4}{npj Quantum
  Information {\bf 4}, 23}~(2018).

\bibitem{BAA+20}
J-H Bae, Paul~M Alsing, Doyeol Ahn, and Warner~A Miller.
\newblock ``{Quantum circuit optimization using quantum Karnaugh map}''.
\newblock \href{https://dx.doi.org/10.1038/s41598-020-72469-7}{Scientific
  reports {\bf 10}, 15651}~(2020).

\bibitem{HS22_TDepth}
Zhenyu Huang and Siwei Sun.
\newblock ``{Synthesizing quantum circuits of AES with lower T-depth and less
  qubits}''.
\newblock In Shweta Agrawal and Dongdai Lin, editors, Advances in Cryptology --
  ASIACRYPT 2022.
\newblock \href{https://dx.doi.org/10.1007/978-3-031-22969-5_21}{Pages
  614--644}.
\newblock Cham~(2022). Springer Nature Switzerland.

\bibitem{AMM14}
Matthew Amy, Dmitri Maslov, and Michele Mosca.
\newblock ``{Polynomial-time T-depth optimization of Clifford+T circuits via
  matroid partitioning}''.
\newblock \href{https://dx.doi.org/10.1109/TCAD.2014.2341953}{IEEE Transactions
  on Computer-Aided Design of Integrated Circuits and Systems {\bf 33},
  1476--1489}~(2014).

\bibitem{bookchuang}
Michael~A. Nielsen and Isaac~L. Chuang.
\newblock ``{Quantum computation and quantum information: 10th anniversary
  edition}''.
\newblock \href{https://dx.doi.org/10.1017/CBO9780511976667}{Cambridge
  University Press}. ~(2010).

\bibitem{maslov21}
Sergey Bravyi, Ruslan Shaydulin, Shaohan Hu, and Dmitri Maslov.
\newblock ``{Clifford circuit optimization with templates and symbolic Pauli
  gates}''.
\newblock \href{https://dx.doi.org/10.22331/q-2021-11-16-580}{{Quantum} {\bf
  5}, 580}~(2021).

\bibitem{bondy2008graph}
John~Adrian Bondy and Uppaluri Siva~Ramachandra Murty.
\newblock ``{Graph theory}''.
\newblock \href{https://dx.doi.org/10.1007/978-1-84628-970-5}{Springer
  Publishing Company, Incorporated}. ~(2008).

\bibitem{KC00}
Valentine Kabanets and Jin-Yi Cai.
\newblock ``{Circuit minimization problem}''.
\newblock In Proceedings of the Thirty-Second Annual ACM Symposium on Theory of
  Computing.
\newblock \href{https://dx.doi.org/10.1145/335305.335314}{Page 73–79}.
\newblock STOC '00New York, NY, USA~(2000). Association for Computing
  Machinery.

\bibitem{AHM+08}
Eric Allender, Lisa Hellerstein, Paul McCabe, Toniann Pitassi, and Michael
  Saks.
\newblock ``{Minimizing disjunctive normal form formulas and $AC^0$ circuits
  given a truth table}''.
\newblock \href{https://dx.doi.org/10.1137/060664537}{SIAM Journal on Computing
  {\bf 38}, 63--84}~(2008).

\bibitem{SIR18}
S~Hirahara, I~Oliveira, and R~Santhanam.
\newblock ``{NP-hardness of minimum circuit size problem for OR-AND-MOD
  circuits}''.
\newblock In 33rd Computational Complexity Conference (CCC 2018).
\newblock \href{https://dx.doi.org/10.4230/LIPIcs.CCC.2018.5}{Volume 102 of
  Leibniz International Proceedings in Informatics, pages 5:1--5:31}.
\newblock Schloss Dagstuhl~(2018).

\bibitem{RBI+20}
Rahul Ilango, Bruno Loff, and Igor~C. Oliveira.
\newblock ``{NP-hardness of circuit minimization for multi-output functions}''.
\newblock In Shubhangi Saraf, editor, 35th Computational Complexity Conference
  (CCC 2020).
\newblock \href{https://dx.doi.org/10.4230/LIPIcs.CCC.2020.22}{Volume 169 of
  Leibniz International Proceedings in Informatics (LIPIcs), pages
  22:1--22:36}.
\newblock Dagstuhl, Germany~(2020). Schloss Dagstuhl -- Leibniz-Zentrum f{\"u}r
  Informatik.

\bibitem{Hir22}
Shuichi Hirahara.
\newblock ``{NP-hardness of learning programs and partial MCSP}''.
\newblock In 2022 IEEE 63rd Annual Symposium on Foundations of Computer Science
  (FOCS).
\newblock \href{https://dx.doi.org/10.1109/FOCS54457.2022.00095}{Pages
  968--979}.
\newblock ~(2022).

\bibitem{braid}
Kunihiro Wasa, Shin Nishio, Koki Suetsugu, Michael Hanks, Ashley Stephens,
  Yu~Yokoi, and Kae Nemoto.
\newblock ``{Hardness of braided quantum circuit optimization in the surface
  code}''.
\newblock \href{https://dx.doi.org/10.1109/TQE.2023.3251358}{IEEE Transactions
  on Quantum Engineering {\bf 4}, 1--7}~(2023).

\bibitem{WA24}
John van~de Wetering and Matt Amy.
\newblock ``{Optimising quantum circuits is generally hard}''~(2024).
\newblock  \href{http://arxiv.org/abs/2310.05958}{arXiv:2310.05958}.

\bibitem{preskill18}
John Preskill.
\newblock ``{Quantum {C}omputing in the {NISQ} era and beyond}''.
\newblock \href{https://dx.doi.org/10.22331/q-2018-08-06-79}{{Quantum} {\bf 2},
  79}~(2018).

\bibitem{amy14}
Matthew Amy, Dmitri Maslov, and Michele Mosca.
\newblock ``{Polynomial-time T-depth optimization of Clifford+T circuits via
  matroid partitioning}''.
\newblock \href{https://dx.doi.org/10.1109/TCAD.2014.2341953}{IEEE Transactions
  on Computer-Aided Design of Integrated Circuits and Systems {\bf 33},
  1476--1489}~(2014).

\bibitem{maslov05}
D.~Maslov, C.~Young, D.M. Miller, and G.W. Dueck.
\newblock ``{Quantum circuit simplification using templates}''.
\newblock In Design, Automation and Test in Europe.
\newblock \href{https://dx.doi.org/10.1109/DATE.2005.249}{Pages 1208--1213 Vol.
  2}.
\newblock ~(2005).

\bibitem{soeken10}
Mathias Soeken, Robert Wille, Gerhard~W. Dueck, and Rolf Drechsler.
\newblock ``{Window optimization of reversible and quantum circuits}''.
\newblock In 13th IEEE Symposium on Design and Diagnostics of Electronic
  Circuits and Systems.
\newblock \href{https://dx.doi.org/10.1109/DDECS.2010.5491754}{Pages 341--345}.
\newblock ~(2010).

\bibitem{MQT}
Robert Wille, Lucas Berent, Tobias Forster, Jagatheesan Kunasaikaran, Kevin
  Mato, Tom Peham, Nils Quetschlich, Damian Rovara, Aaron Sander, Ludwig
  Schmid, Daniel Schönberger, Yannick Stade, and Lukas Burgholzer.
\newblock ``{The MQT handbook : A summary of design automation tools and
  software for quantum computing}''.
\newblock In 2024 IEEE International Conference on Quantum Software (QSW).
\newblock \href{https://dx.doi.org/10.1109/QSW62656.2024.00013}{Pages 1--8}.
\newblock ~(2024).

\bibitem{qiskit2024}
Ali Javadi-Abhari, Matthew Treinish, Kevin Krsulich, Christopher~J. Wood, Jake
  Lishman, Julien Gacon, Simon Martiel, Paul~D. Nation, Lev~S. Bishop,
  Andrew~W. Cross, Blake~R. Johnson, and Jay~M. Gambetta.
\newblock ``Quantum computing with {Q}iskit''~(2024).
\newblock  \href{http://arxiv.org/abs/2405.08810}{arXiv:2405.08810}.

\bibitem{MZ22}
Dmitri Maslov and Ben Zindorf.
\newblock ``{Depth optimization of CZ, CNOT, and Clifford circuits}''.
\newblock \href{https://dx.doi.org/10.1109/TQE.2022.3180900}{IEEE Transactions
  on Quantum Engineering {\bf 3}, 1--8}~(2022).

\bibitem{mulGNB}
Brittanney Amento, Martin R\"{o}tteler, and Rainer Steinwandt.
\newblock ``{Quantum binary field inversion: improved circuit depth via choice
  of basis representation}''.
\newblock \href{https://dx.doi.org/10.26421/qic13.1-2-7}{Quantum Info. Comput.
  {\bf 13}, 116–134}~(2013).

\bibitem{HOH21}
Rebekah Herrman, James Ostrowski, Travis~S. Humble, and George Siopsis.
\newblock ``{Lower bounds on circuit depth of the quantum approximate
  optimization algorithm}''.
\newblock \href{https://dx.doi.org/10.1007/s11128-021-03001-7}{Quantum
  Information Processing {\bf 20}, 59}~(2021).

\bibitem{bookBergeHypergraph}
Claude Berge.
\newblock ``Hypergraphs: combinatorics of finite sets''.
\newblock \href{https://dx.doi.org/10.1016/s0924-6509(08)x7007-2}{Volume~45}.
\newblock Elsevier. ~(1984).

\bibitem{JGM19}
Andrew Jena, Scott Genin, and Michele Mosca.
\newblock ``{Pauli partitioning with respect to gate sets}''~(2019).
\newblock  \href{http://arxiv.org/abs/1907.07859}{arXiv:1907.07859}.

\bibitem{VYI20}
Vladyslav Verteletskyi, Tzu-Ching Yen, and Artur~F. Izmaylov.
\newblock ``{Measurement optimization in the variational quantum eigensolver
  using a minimum clique cover}''.
\newblock \href{https://dx.doi.org/10.1063/1.5141458}{The Journal of Chemical
  Physics {\bf 152}, 124114}~(2020).

\bibitem{HI20}
Ikko Hamamura and Takashi Imamichi.
\newblock ``{Efficient evaluation of quantum observables using entangled
  measurements}''.
\newblock \href{https://dx.doi.org/10.1038/s41534-020-0284-2}{npj Quantum
  Information {\bf 6}, 56}~(2020).

\bibitem{GAD20}
Pranav Gokhale, Olivia Angiuli, Yongshan Ding, Kaiwen Gui, Teague Tomesh,
  Martin Suchara, Margaret Martonosi, and Frederic~T. Chong.
\newblock ``{$O(N^3)$ measurement cost for variational quantum eigensolver on
  molecular hamiltonians}''.
\newblock \href{https://dx.doi.org/10.1109/TQE.2020.3035814}{IEEE Transactions
  on Quantum Engineering {\bf 1}, 1--24}~(2020).

\bibitem{VQE}
Alberto Peruzzo, Jarrod McClean, Peter Shadbolt, Man-Hong Yung, Xiao-Qi Zhou,
  Peter~J. Love, Al{\'a}n Aspuru-Guzik, and Jeremy~L. O’Brien.
\newblock ``{A variational eigenvalue solver on a photonic quantum
  processor}''.
\newblock \href{https://dx.doi.org/10.1038/ncomms5213}{Nature Communications
  {\bf 5}, 4213}~(2014).

\bibitem{AJS24}
Muhammad Asaduzzaman, Raghav~G. Jha, and Bharath Sambasivam.
\newblock ``{Sachdev-Ye-Kitaev model on a noisy quantum computer}''.
\newblock \href{https://dx.doi.org/10.1103/PhysRevD.109.105002}{Phys. Rev. D
  {\bf 109}, 105002}~(2024).

\bibitem{WP67}
D.~J.~A. Welsh and M.~B. Powell.
\newblock ``{An upper bound for the chromatic number of a graph and its
  application to timetabling problems}''.
\newblock \href{https://dx.doi.org/10.1093/comjnl/10.1.85}{The Computer Journal
  {\bf 10}, 85--86}~(1967).

\bibitem{ColDSatur}
Daniel Br\'{e}laz.
\newblock ``{New methods to color the vertices of a graph}''.
\newblock \href{https://dx.doi.org/10.1145/359094.359101}{Commun. ACM {\bf 22},
  251–256}~(1979).

\bibitem{Lei79}
Frank~Thomson Leighton.
\newblock ``{A graph coloring algorithm for large scheduling problems}''.
\newblock \href{https://dx.doi.org/10.6028/jres.084.024}{Journal of research of
  the national bureau of standards {\bf 84}, 489--506}~(1979).

\bibitem{AS06}
Hussein Al-Omari and Khair~Eddin Sabri.
\newblock ``{New graph coloring algorithms}''.
\newblock \href{https://dx.doi.org/10.3844/jmssp.2006.439.441}{American Journal
  of Mathematics and Statistics {\bf 2}, 739--741}~(2006).

\bibitem{DSaturBenchmark1}
Murat Aslan and Nurdan~Akhan Baykan.
\newblock ``A performance comparison of graph coloring algorithms''.
\newblock \href{https://dx.doi.org/10.18201/ijisae.273053}{International
  Journal of Intelligent Systems and Applications in Engineering {\bf 4},
  1--7}~(2016).

\bibitem{DSaturBenchmark2}
Isabelle Devarenne, Hakim Mabed, and Alexandre Caminada.
\newblock ``Intelligent neighborhood exploration in local search heuristics''.
\newblock In 2006 18th IEEE International Conference on Tools with Artificial
  Intelligence (ICTAI'06).
\newblock \href{https://dx.doi.org/10.1109/ICTAI.2006.68}{Pages 144--150}.
\newblock ~(2006).

\bibitem{backtracting}
J.~Randall Brown.
\newblock ``Chromatic scheduling and the chromatic number problem''.
\newblock \href{https://dx.doi.org/10.1287/mnsc.19.4.456}{Management Science
  {\bf 19}, 456--463}~(1972).

\bibitem{RIG99}
Micha{\l} Karo{\'n}ski, Edward~R Scheinerman, and Karen~B Singer-Cohen.
\newblock ``{On random intersection graphs: The subgraph problem}''.
\newblock \href{https://dx.doi.org/10.1017/S0963548398003459}{Combinatorics,
  Probability and Computing {\bf 8}, 131–159}~(1999).

\bibitem{BTU09}
Michael Behrisch, Anusch Taraz, and Michael Ueckerdt.
\newblock ``{Coloring random intersection graphs and complex Networks}''.
\newblock \href{https://dx.doi.org/10.1137/050647153}{SIAM Journal on Discrete
  Mathematics {\bf 23}, 288--299}~(2009).

\bibitem{KR15}
Valentas Kurauskas and Katarzyna Rybarczyk.
\newblock ``{On the chromatic index of random uniform hypergraphs}''.
\newblock \href{https://dx.doi.org/10.1137/130942292}{SIAM Journal on Discrete
  Mathematics {\bf 29}, 541--558}~(2015).

\bibitem{Ryb17}
Katarzyna Rybarczyk.
\newblock ``{The chromatic number of random intersection graphs}''.
\newblock
  \href{https://dx.doi.org/https://doi.org/10.7151/dmgt.1955}{Discussiones
  Mathematicae Graph Theory {\bf 37}, 465--476}~(2017).

\bibitem{mulghost}
Brittanney Amento, Martin R\"{o}tteler, and Rainer Steinwandt.
\newblock ``{Efficient quantum circuits for binary elliptic curve arithmetic:
  reducing T-gate complexity}''.
\newblock \href{https://dx.doi.org/10.26421/qic13.7-8-5}{Quantum Info. Comput.
  {\bf 13}, 631–644}~(2013).

\bibitem{RS14}
Martin R\"{o}tteler and Rainer Steinwandt.
\newblock ``{A quantum circuit to find discrete logarithms on ordinary binary
  elliptic curves in depth $O(\log^2 n)$}''.
\newblock \href{https://dx.doi.org/10.26421/QIC14.9-10-11}{Quantum Info.
  Comput. {\bf 14}, 888–900}~(2014).

\bibitem{mulKepley}
Shane Kepley and Rainer Steinwandt.
\newblock ``{Quantum circuits for $\mathbb{F}_{2^{n}}$-multiplication with
  subquadratic gate count}''.
\newblock \href{https://dx.doi.org/10.1007/s11128-015-0993-1}{Quantum
  Information Processing {\bf 14}, 2373--2386}~(2015).

\bibitem{mulSpaceEfficient}
Iggy Van~Hoof.
\newblock ``{Space-efficient quantum multiplication of polynomials for binary
  finite fields with sub-quadratic Toffoli gate count}''.
\newblock \href{https://dx.doi.org/10.26421/qic20.9-10-1}{Quantum Info. Comput.
  {\bf 20}, 721--735}~(2020).

\bibitem{mulCF}
Qing-bin Luo, Guo-wu Yang, Xiao-yu Li, and Qiang Li.
\newblock ``{Quantum reversible circuits for $\mathrm{GF}(2^{8})$
  multiplicative inverse}''.
\newblock \href{https://dx.doi.org/10.1140/epjqt/s40507-022-00144-z}{EPJ
  Quantum Technology {\bf 9}, 24}~(2022).

\bibitem{pyqubo}
Mashiyat Zaman, Kotaro Tanahashi, and Shu Tanaka.
\newblock ``{PyQUBO: python library for mapping combinatorial optimization
  problems to QUBO form}''.
\newblock \href{https://dx.doi.org/10.1109/tc.2021.3063618}{IEEE Transactions
  on Computers {\bf 71}, 838--850}~(2022).

\bibitem{RG17}
Lidia Ruiz-Perez and Juan~Carlos Garcia-Escartin.
\newblock ``{Quantum arithmetic with the quantum Fourier transform}''.
\newblock \href{https://dx.doi.org/10.1007/s11128-017-1603-1}{Quantum
  Information Processing {\bf 16}, 152}~(2017).

\bibitem{Bol98}
B.~Bollob\'{a}s.
\newblock ``{The chromatic number of random graphs}''.
\newblock \href{https://dx.doi.org/10.1007/BF02122551}{Combinatorica {\bf 8},
  49--55}~(1988).

\bibitem{zalka}
Christof Zalka.
\newblock ``{Grover's quantum searching algorithm is optimal}''.
\newblock \href{https://dx.doi.org/10.1103/PhysRevA.60.2746}{Phys. Rev. A {\bf
  60}, 2746--2751}~(1999).

\bibitem{PLU_SODA}
Jiaqing Jiang, Xiaoming Sun, Shang-Hua Teng, Bujiao Wu, Kewen Wu, and Jialin
  Zhang.
\newblock ``{Optimal space-depth trade-off of CNOT circuits in quantum logic
  synthesis}''.
\newblock In Proceedings of the Thirty-First Annual ACM-SIAM Symposium on
  Discrete Algorithms.
\newblock \href{https://dx.doi.org/10.1137/1.9781611975994.13}{Page 213–229}.
\newblock SODA '20USA~(2020). Society for Industrial and Applied Mathematics.

\bibitem{PolyTimePerfectColoring}
M.~Gr{\"o}tschel, L.~Lov{\'a}sz, and A.~Schrijver.
\newblock ``{Polynomial algorithms for perfect graphs}''.
\newblock In C.~Berge and V.~Chv{\'a}tal, editors, Topics on Perfect Graphs.
\newblock \href{https://dx.doi.org/10.1016/S0304-0208(08)72943-8}{Volume~88 of
  North-Holland Mathematics Studies, pages 325--356}.
\newblock North-Holland~(1984).

\end{thebibliography}

\onecolumn
\appendix

\let\oldv=\v
\renewcommand{\v}[1]{{\bm #1}}
\section{Reduction from graph coloring to gate ordering}\label{sec:reduction-other-way}
The graph coloring problem can be reduced to the gate ordering. 
An algorithm that takes as input a graph and outputs a circuit is required for the reduction, but since parallelizability of gates is dependent on the underlying hardware, it is difficult to describe the algorithm in an abstract manner.
Therefore, we fix the condition for two gates to be parallelizable by taking Assumption\;\ref{assump:parallelizability} from the main text, but a similar algorithm can be developed for different conditions in general.

Let $C\!Z_{\mathcal{C}}$ denote a multi-controlled phase shift gate (by an arbitrary angle $\theta$), where $\mathcal{C}$ is the set of referenced qubits.
For instance, $C\!Z_{\{q_{1},q_{2}\}}$ gate endows the quantum state with phase $e^{i\theta}$ when $q_{1}=q_{2}=1$.

\begin{algorithm}[H]
	\small
	\caption{$\mathsf{ConstructCircuit}$}
	\begin{algorithmic}[1]
		\Require $G = (V, E)$
		\Ensure $\v v = (v_{1}, \ldots, v_{|V|})$
		\algrule
		\State $t \gets 1$
		\For{$i = 1$ \textbf{to} $|V|$}
			\State $\mathcal{C}_{i} \gets \{q_{1}, \ldots, q_{t}\}$, $\mathcal{T} \gets \{\}$
			\For{$j = 1$ \textbf{to} $i-1$}
				\If{$e_{ij} \not\in E$}
					\State $\mathcal{C}_{i} \gets \mathcal{C}_{i} \backslash \mathcal{C}_{j}$
				\EndIf
			\EndFor
			\For{$j = 1$ \textbf{to} $i-1$}
				\If{$e_{ij} \in E$ \textbf{and} $\mathcal{C}_{i} \cap \mathcal{C}_{j} = \{\}$}
					\For{$q \in \mathcal{T}$}
						\State $p \gets \mathsf{True}$
						\For{$k=1$ \textbf{to} $j-1$}
							\If{$e_{kj} \not\in E$ \textbf{and} $q \in \mathcal{C}_{k}$}
								\State $p \gets \mathsf{False}$
								\State \textbf{break}
							\EndIf
						\EndFor
						\If{$p = \mathsf{True}$}
							\State $\mathcal{C}_{i} \gets \mathcal{C}_{i} \cup \{q\}$, $\mathcal{C}_{j} \gets \mathcal{C}_{j} \cup \{q\}$
							\State \textbf{continue}
						\EndIf
					\EndFor
					\If{$\mathcal{C}_{j} \cap \mathcal{T} = \{\}$}
						\State $t \gets t + 1$
						\State $\mathcal{C}_{i} \gets \mathcal{C}_{i} \cup \{q_{t}\}$, $\mathcal{C}_{j} \gets \mathcal{C}_{j} \cup \{q_{t}\}$
						\State $\mathcal{T} \gets \mathcal{T} \cup \{t\}$
					\EndIf
				\EndIf
			\EndFor
			\If{$\mathcal{C}_{i} = \{\}$}
				\State $t \gets t+1$
				\State $\mathcal{C}_{i} \gets \mathcal{C}_{i} \cup \{q_{t}\}$
			\EndIf
		\EndFor
		\For{$i=1$ \textbf{to} $|V|$}
			\State $v_{i} \gets C\!Z_{\mathcal{C}_{i}}$
		\EndFor
		\State \Return $\v v = (v_{1}, \ldots, v_{|V|})$
	\end{algorithmic}
	\label{alg:graph_to_circuit_app}
\end{algorithm}

\begin{figure}[H]
	\centering
	\includegraphics[width=0.75\textwidth]{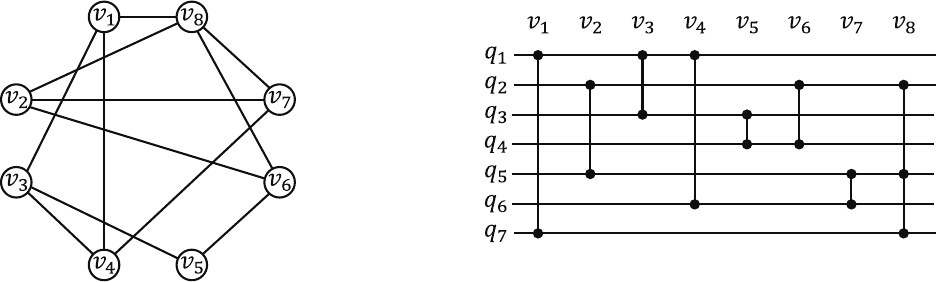}
	\caption{An input graph (left) and the output circuit (right) of Algorithm~\ref{alg:graph_to_circuit_app}. Each gate in the circuit is a phase shift gate.}
	\label{fig:exmaple_algo4}
\end{figure}

\noindent Algorithm\;\ref{alg:graph_to_circuit_app} constructs a circuit such that any two connected vertices (gates) share at least one qubit as a common reference.
The resulting circuit consists of $|V|$ commuting gates and no more than $|V|^{2}/2$ qubits.
(We have observed that the number of qubits tends to get smaller as the edge density approaches either zero or 1.)
Figure\;\ref{fig:exmaple_algo4} illustrates how the circuit is composed by Algorithm\;\ref{alg:graph_to_circuit_app}.
Note that for given graph $G$, we have $G = \mathsf{ConstructGraph} (\mathsf{ConstructCircuit}(G))$, where $\mathsf{ConstructGraph}$ is Algorithm\;\ref{alg:construct_graph}.

Given a graph $G$, we construct a commuting circuit $\v v_{G}$ corresponding to $G$ by Algorithm~\ref{alg:graph_to_circuit_app}.
Let $\mathcal{V}$ be the multiset of gates in $\v v_{G}$, let $A_{\mathcal{V}}$ be the set of all permutations of gates in $\mathcal{V}$, and let $B_{\mathcal{V}}$ be the set of all colored graphs of $G$.
Since $G = \mathsf{ConstructGraph} (\mathsf{ConstructCircuit}(G))$, Claim\;\ref{cla:functions} still holds and we have two functions $f$ and $g$ as stated in the claim.
Then we have the following proposition:

\begin{myProposition}
	If $m_{\rm depth}(\v a) = \displaystyle\min_{\v a' \in A_{\mathcal{V}}} m_{\rm depth}(\v a')$ holds, and then it is also true that $m_{\rm color}\big( g(\v a) \big) = \displaystyle\min_{b' \in B_{\mathcal{V}}} \big( m_{\rm color}(b') \big)$.
	\label{prop:reduction reverse}
\end{myProposition}
\begin{proof}
	We use the proof by contradiction.
	Let ${\v a} \in A_{\mathcal{V}}$ be the circuit such that
	\begin{align}\label{eq:reduction-2}
		m_{\rm depth}(\v a) = \min_{\v a' \in A_{\mathcal{V}}} \big(m_{\rm depth}(\v a') \big) \enspace.
	\end{align}
	Now suppose that $m_{\rm color}(g(\v a)) > \min_{b' \in B_{\mathcal{V}}} \big( m_{\rm color}(b') \big)$, then there exists $b \in B_{\mathcal{V}}$ such that $m_{\rm color}(b) < m_{\rm color}(g(\v a))$, $b \neq g(\v a)$.
	
	Due to the properties of $f$ and $g$, we have $m_{\rm depth}(f(b)) \le m_{\rm color}(b)$ and $m_{\rm color}(g(\v a)) = m_{\rm depth}(\v a)$.
	In other words, there exists $g(\v a) \in B_{\cal V}$ satisfying
	\begin{align*}
		m_{\rm depth}(f(b)) \le m_{\rm color}(b) < m_{\rm color}(g(\v a)) = m_{\rm depth}(\v a) \enspace,
	\end{align*}
	which contradicts to Eq.\;(\ref{eq:reduction-2}).
\end{proof}


\section{On remark~\ref{remark:prop_Algo2}}\label{sec:remark6}
\begin{figure}[htbp]
	\centering
	\includegraphics[width=0.9\textwidth]{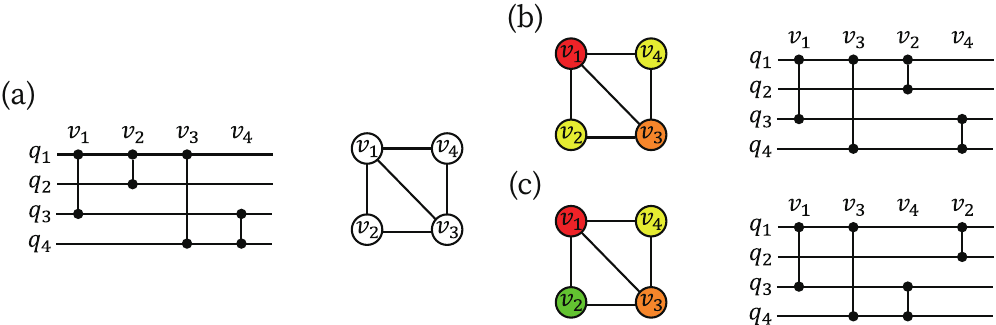}
	\caption{A commuting circuit and corresponding graph are shown in (a), and its colored graphs and resulting circuits are shown in (b) and (c).}
	\label{fig:ex_remark6}
\end{figure}

Case\;(a) in Remark~\ref{remark:prop_Algo2} can only occurs when the coloring algorithm produces an approximate solution, not the true optimal one.
Figure~\ref{fig:ex_remark6}\;(a) illustrates a commuting circuit and its corresponding graph.
Having Assumption\;\ref{assump:parallelizability}, the depth of the circuit is four.
In Figure~\ref{fig:ex_remark6}\;(b), the graph is colored with the minimum number of colors, yielding the optimal solution.
Consequently, the number of colors coincides with the depth of the resulting circuit.
By contrast, Figure~\ref{fig:ex_remark6}\;(c) shows an approximate solution for the coloring.
However, the depth of the resulting circuit is three since $v_2$ and $v_4$ happen to be simultaneously executable.


\section{Coloring within Qiskit framework}\label{sec:Qiskit}
Since the proposed method and Qiskit transpiler's optimization are not directly comparable, we let the Qiskit compiler optimize commuting circuits, with and without the coloring method applied beforehand, into Clifford+$T$ circuits.
By measuring the differences in two cases (with and without coloring applied), we assess how effectively the coloring enhances the compiler.

\begin{figure}[htbp]
	\centering
	\includegraphics[width=0.8\textwidth]{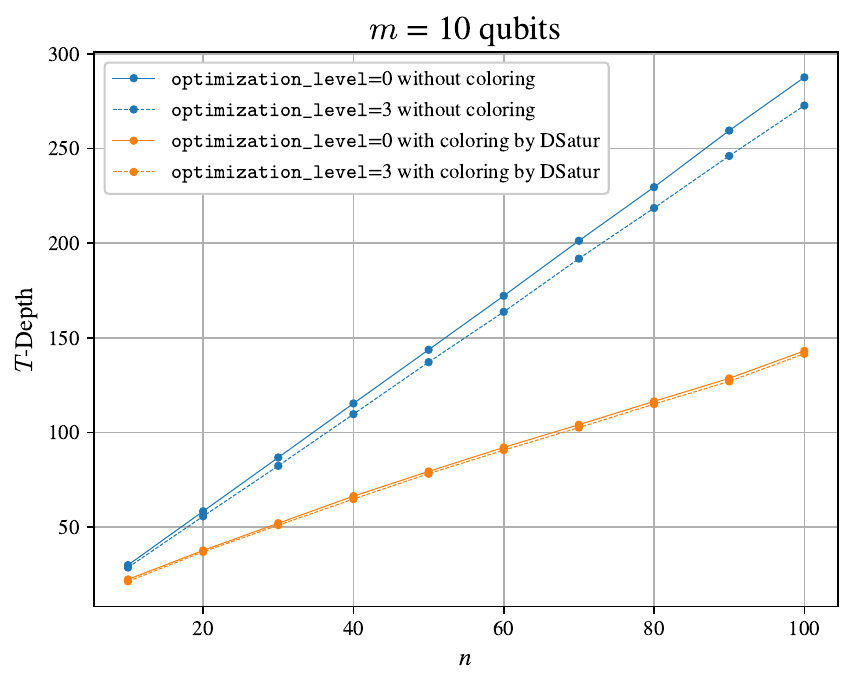}
	\caption{Optimization by Qiskit transpiler with and without the coloring method applied for randomly generated commuting circuits with $m = 10$ qubits and $n \in \{10,20,...,100\}$ gates.
		Markers with solid and dashed lines are output by Qiskit with \texttt{optimization\_level}$\;=0$ and \texttt{optimization\_level}$\;=3$, respectively.}
	\label{fig:experiemnt_qiskit}
\end{figure}

We randomly generate commuting circuits with $m = 10$ qubits and $n \in \{10,20,...,100\}$ CCZ gates.
Unlike the experiments in the main text, we fixed the size of the gates because when decomposing into a Clifford+$T$ circuit, imposing the same degree of parallelization on gates of different sizes (e.g., CCZ and CCCZ) is not well aligned with the goal of the experimental setup.
The key Qiskit transpiler settings are as follows: \texttt{OptimizationMetric} is set to \texttt{COUNT\_T} and \texttt{optimization\_level} is set to 3.
The results are shown in Figure\;\ref{fig:experiemnt_qiskit}.

\let\v=\oldv

\end{document}